\documentclass[a4paper,11pt]{article}
\pdfoutput=1 

\usepackage{jheppub} 

\usepackage[T1]{fontenc} 

\usepackage{ifthen}
\usepackage{tikz}

\usepackage{color}

\newcommand{\GeV}{{\rm GeV}}

\newcommand{\vev}[1]{\langle #1 \rangle}

\newcommand{\lrb}[1]{\left( #1 \right)}

\newcommand{\eq}[2]{#1^{\, \rm eq}_{#2}}

\title{Dark matter freeze-in from semi-production}

\author{Andrzej Hryczuk}
\author{{\rm and} Maxim Laletin}

\affiliation{National Centre for Nuclear Research,\\ Pasteura 7, 02-093 Warsaw, Poland}

\emailAdd{andrzej.hryczuk@ncbj.gov.pl}
\emailAdd{maxim.laletin@ncbj.gov.pl}

\abstract{We study a novel dark matter production mechanism based on the freeze-in through \textit{semi-production}, \textit{i.e.} the inverse semi-annihilation processes. A peculiar feature of this scenario is that the production rate is suppressed by a small initial abundance of dark matter and consequently creating the observed abundance requires much larger coupling values than for the usual freeze-in. We provide a concrete example model exhibiting such production mechanism and study it in detail, extending the standard formalism to include the evolution of dark matter temperature alongside its number density and discuss the importance of this improved treatment. Finally, we confront the relic density constraint with the limits and prospects for the dark matter indirect detection searches. We show that, even if it was never in full thermal equilibrium in the early Universe, dark matter could, nevertheless, have strong enough present-day annihilation cross section to lead to observable signals.
}

\begin{document} 
\maketitle
\flushbottom

\section{Introduction}

Although the nature of dark matter (DM) remains an open question, the measurement of its present-day energy density  constitutes a robust positive signal of physics beyond the Standard Model (SM). A successful explanation of the properties of DM, thus, by necessity needs to provide a production mechanism that can fit the observed value of $\Omega_{\text{DM}} h^2 = 0.12\pm 0.0012$~\cite{Aghanim:2018eyx}. Even though this is quite a constrictive requirement within any given particle physics model, numerous ways to explain the current abundance are known when being completely agnostic about the nature of DM. Among these of special interest are mechanisms that rely solely on interactions of the dark sector states with the SM ones from the thermal plasma and do not carry any dependence on the details of the initial conditions at the time of reheating. In particular, the thermal freeze-out from the 
primordial plasma~\cite{Lee:1977ua}, where the DM at some time in its evolution was in chemical equilibrium, and freeze-in~\cite{Hall:2009bx,Chu:2011be}, where its abundance was always much lower than in equilibrium and DM was gradually produced from negligible starting population.

While the freeze-out paradigm often leads to DM candidates that have appreciable signals in direct, indirect and collider searches and is subject to increasingly far-reaching constraints (for a review see \textit{e.g.} \cite{Arcadi:2017kky}) the freeze-in naturally predicts the DM particles to be much more weakly interacting and more challenging to detect (see \textit{e.g.} \cite{Bernal:2017kxu}). In particular, across the board expectation for stable DM produced through freeze-in is to provide null signal in the indirect detection searches (see \cite{Heikinheimo:2014xza,Heikinheimo:2018duk} for exceptions).

The specific processes that lead to the freeze-in production vary depending on the model, but the most well-studied realizations are coming from the decays or $2\rightarrow 2$  pair-production processes from the states that are in equilibrium with the SM plasma (see again \cite{Bernal:2017kxu} for a comprehensive review). Analogous processes play a vital role also in the majority of models based on the freeze-out mechanism. Interestingly, in the latter context a different process has become of growing interest in the recent years, namely the semi-annihilation \cite{DEramo:2010keq}, \textit{i.e.} the interaction of the type $\chi\chi \rightarrow \chi\phi$, where $\chi$ represents the DM particle, while $\phi$ is either a SM state or a mediator. Semi-annihilations are present in a number of realistic models (see \textit{e.g.}~\cite{Belanger:2012zr,Belanger:2012vp,Rodejohann:2015lca,Cai:2015zza}) and not only are they subject to weakened direct detection limits, but also lead to peculiar phenomenology concerning the indirect detection \cite{Cai:2018imb} and the core-cusp cosmological problem~\cite{Kamada:2017gfc,Kamada:2018hte,Chu:2018nki}.

In this work we study a novel freeze-in production mechanism based on the inverse semi-annihilation processes, which we will refer to as \textit{semi-production}. A peculiar feature of this mechanism is that the production rate is suppressed not only by small couplings, but also by a small initial abundance of DM such that it  vanishes in the limit when no DM particles are present in the plasma. Hence, creating the observed abundance requires much larger coupling values than in the usual freeze-in scenario. Consequently, even if DM was never in full thermal equilibrium (with the SM plasma or within the dark sector) in the early Universe, it can still have strong enough present-day annihilation cross section to be a source of observable signals in the indirect searches.

An interesting effect that arises in the study of the proposed mechanism is that the semi-production rate can strongly depend not only on the number density of DM particles at any given time, but also on their energy. Especially when the semi-annihilating partner is lighter than the DM particle, which causes a kinematical suppression of the interactions involving low energy tails of the momentum distribution. Therefore, in order to provide a reliable calculation of the resulting abundance we solve the coupled system of Boltzmann equations for both the number density and the temperature of DM.

This paper is organized as follows. We start in Sec.~\ref{sec:illustration} with the basic illustration of the mechanism. In Sec.~\ref{sec:model} we describe the specifics of the model chosen to demonstrate the properties of the proposed scenario. Sec.~\ref{sec:RD} discusses the  formalism suitable for determining the DM relic density in this setup. Finally, in Sec.~\ref{sec:scan} we present results of the scan in the parameter space of the model and discuss phenomenology, while Sec.~\ref{sec:conclusions} concludes.

\section{Illustration of the mechanism}
\label{sec:illustration}

To introduce the mechanism let us begin with a  simple example. Consider a model extending the SM by a real scalar $\phi$, which we will refer to as the mediator, and a complex scalar $\chi$ charged under a $Z_3$ symmetry, constituting the DM. Assume that the interaction Lagrangian is given by
\begin{equation}
\label{eq:Ltoy}
\mathcal{L}_{int} = \mathcal{L}_{SM}  +\mathcal{L}_{\phi-SM} + \frac{\lambda}{3!} \phi\left( \chi^3 + (\chi^*)^3\right),
\end{equation}
for now neglecting all the other possible interaction terms. Furthermore, let us assume that the mediator is tightly coupled to the SM plasma keeping it both in thermal and chemical equilibrium for all the time that number changing processes for $\chi$ are active. Finally, let us focus on the region where $m_\phi < 3m_\chi$, such that the $\phi \rightarrow \chi\chi\chi$ decay is kinematically forbidden and that $\lambda \ll 1$ such that $\chi$ never thermalizes. 

In such an example toy model, the DM relic density is build up by the freeze-in mechanism due to the $\phi\chi \rightarrow \chi\chi$ \textit{semi-production} process. However, in contrast to the usual freeze-in, the rate for this process vanishes in the limit of vanishing number density of $\chi$. In other words, in order for it to take place there needs to be some initial population of $\chi$'s present. This could be \textit{e.g.} a direct remnant of the reheating process~\cite{Takahashi:2007tz}, effect of the gravitational production \cite{Garny:2015sjg, Mambrini:2021zpp}, ultraviolet freeze-in \cite{Moroi:1993mb,Bolz:2000fu}, forbidden freeze-in \cite{Darme:2019wpd} (see also~\cite{Biondini:2020ric}) or decay of the false vacua~\cite{Asadi:2021pwo}. 
In fact, a general renormalizable Lagrangian that accommodates semi-annihilation also typically allows for the pair-production process (see Sec.~\ref{sec:model}). Such a process can, thus, naturally accompany semi-annihilation and explain the genesis of the initial dark matter population.

Irrespectively of the details of the mechanism giving rise to this initial population of $\chi$, as long as it happens at some early time corresponding to the temperature of the Universe being $T_{in}$, the yield $Y_\chi(T_{in}) \equiv Y^{in}_\chi \neq 0$, the semi-production will be active. However, its effect is going to be suppressed by a very low number density of $\chi$'s, leading to the requirement for the coupling $\lambda$ to be higher than for the usual $2\rightarrow 2$ freeze-in.

The evolution of the DM yield $Y_\chi=n_\chi/s$ is then given by the Boltzmann equation which can be written as (see Sec.~\ref{sec:BE} for more details):
\begin{equation}
\label{eq:BE0}
\frac{dY}{dx}=\lambda^2\gamma_{\phi\chi\rightarrow\chi\chi}(x,x_\chi) Y \, ,
\end{equation}
where the function $\gamma_{\phi\chi\rightarrow\chi\chi}$ in general depends not only on the SM temperature parameter $x=m_\chi/T$, but also on the form of the distribution function of $\chi$. In particular, making a simplifying assumption that its shape traces the equilibrium one, but allowing for a different temperature $T_\chi$, introducing $x_\chi=m_\chi/T_\chi$ one has 
\begin{equation}
\gamma_{\phi\chi\rightarrow\chi\chi}(x,x_\chi) = \frac{1}{x s \tilde H} \int \frac{d^3k}{(2\pi)^3} \frac{d^3p}{(2\pi)^3} \sigma_{\phi\chi\rightarrow\chi\chi}v\, f_\phi(k,T) f_\chi(p,T_\chi),
\end{equation}
where $\tilde H\equiv H/\left[1+ 1/3d(\log h_{\rm eff})/d(\log T)\right]$, with $H$ being the Hubble rate, $h_{\rm eff}$ is the number of entropy degrees of freedom and $s$ is the entropy density.

If the function $x_\chi(x)$ is known \textit{a priori}, the number density can be calculated simply by integrating \eqref{eq:BE0} giving a formal solution:
\begin{eqnarray}
Y(x_0) =  Y_{in} \exp\left(\lambda^2{\int_{x_{in}}^{x_0} dx\, \gamma_{\phi\chi\rightarrow\chi\chi}(x,x_\chi(x))}\right).
\end{eqnarray}
Inverting this relation one gets the prediction for the coupling $\lambda$ that results in the observed relic abundance to be
\begin{equation}
\label{eq:lambda2}
\lambda^2 =\left[ \int_{x_{in}}^{x_0} dx\, \gamma_{\phi\chi\rightarrow\chi\chi}(x,x_\chi(x))  \right]^{-1} \log\frac{Y_0}{Y_{in}}\, ,
\end{equation}
where $Y_0 = 0.12 (\rho_{c}/(m_\chi s_0))$ with $\rho_c$ and $s_0$ being the critical density and entropy density today respectively.

\begin{figure}[t]
\centering
\includegraphics[scale=0.505]{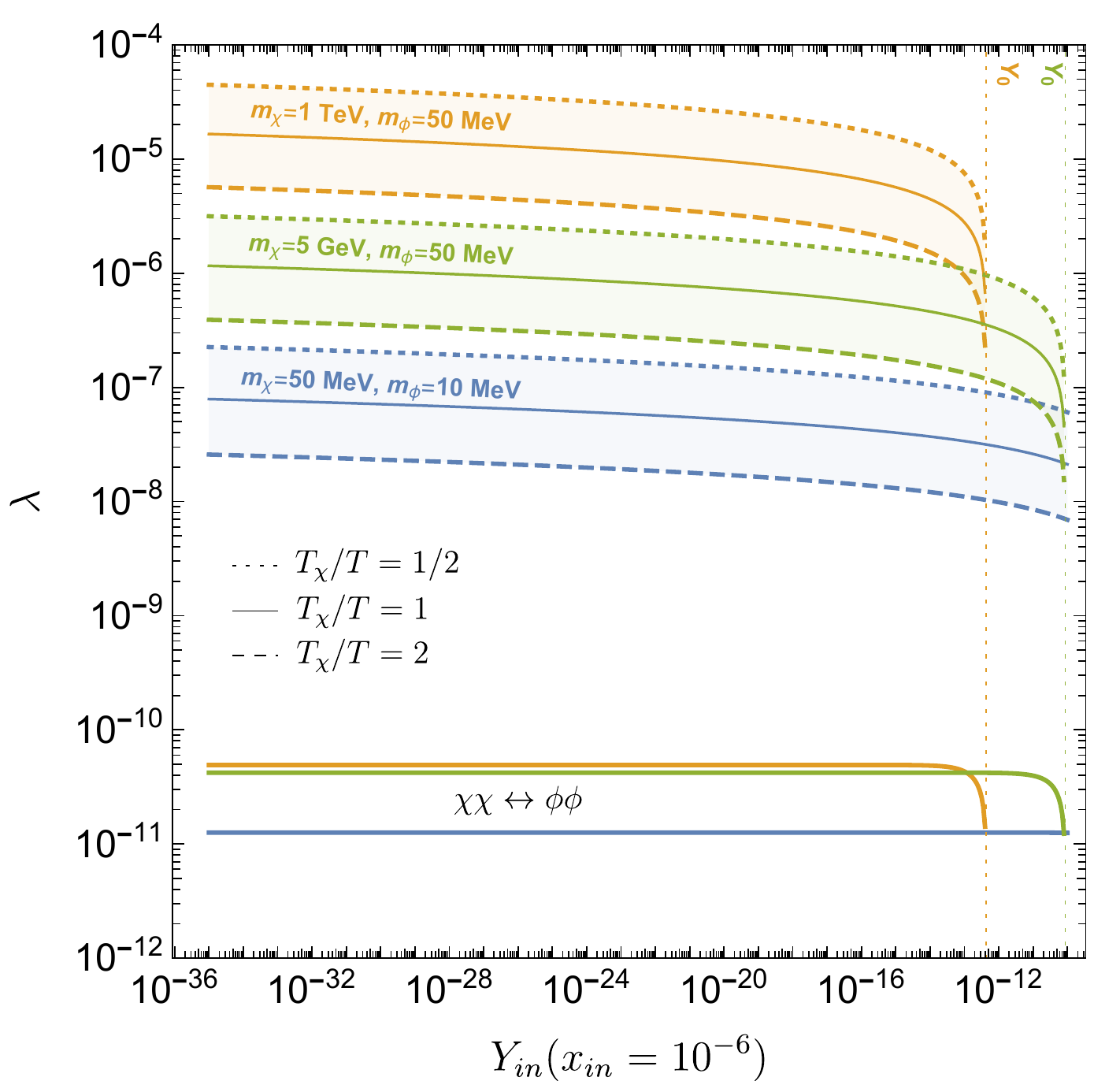}
\hspace{0.12cm}
\includegraphics[scale=0.51]{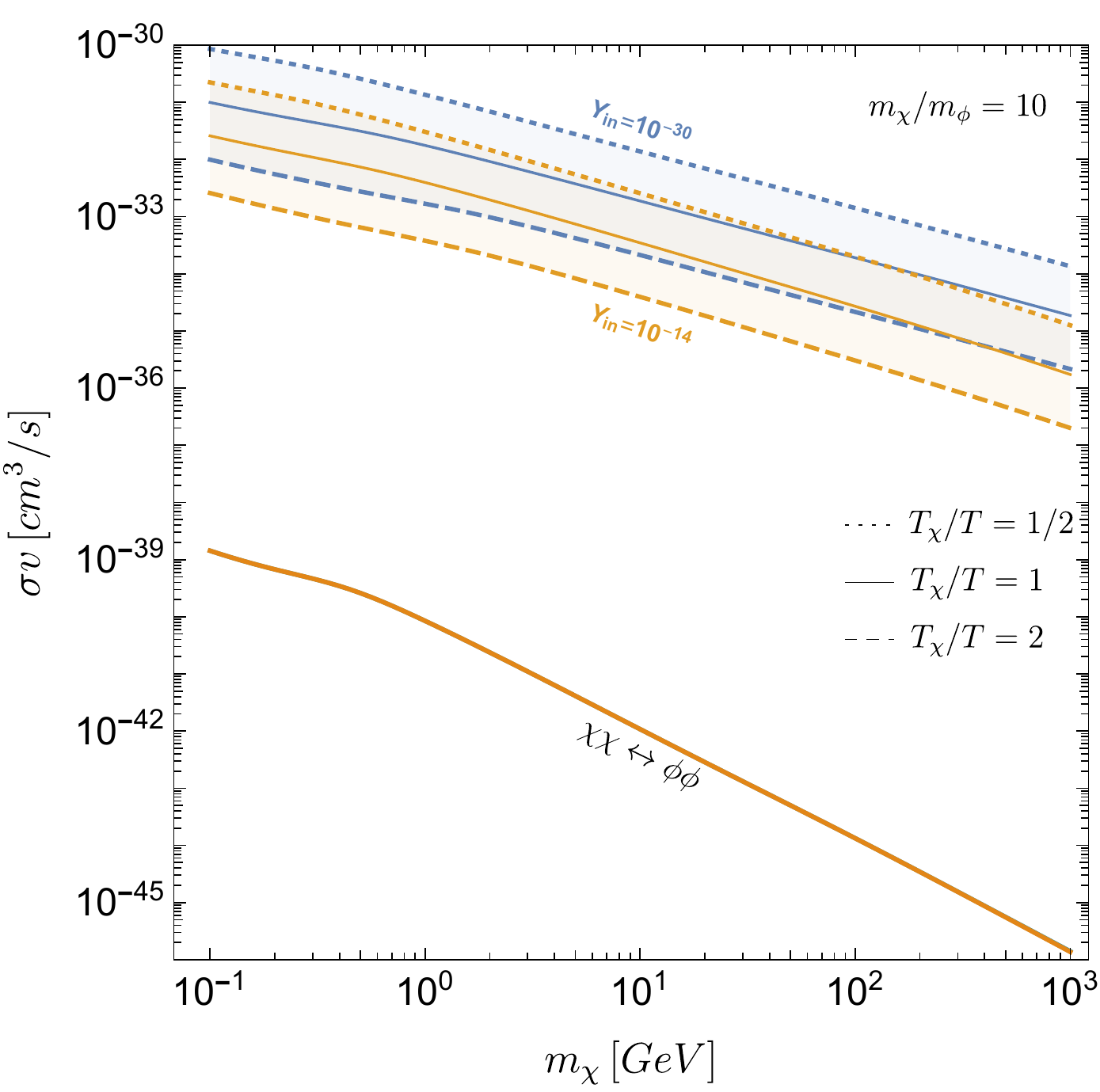}
\caption{The value of the coupling $\lambda$ \textit{(left)} and the present day annihilation cross section \textit{(right)} giving $\Omega h^2= \Omega_{\rm DM} h^2$ through the semi-production mechanism for different choices of masses and fixed $T_\chi/T$. The solid lines give the corresponding values if the process was  $\phi\phi \rightarrow \chi\chi$ instead. 
}
\label{fig:illustration}
\end{figure}

To illustrate the difference with respect to the usual freeze-in and highlight the importance of the evolution of $T_\chi$ alongside its number density in the left panel of Fig.~\ref{fig:illustration} we show the value of the coupling \eqref{eq:lambda2} resulting in the observed relic abundance as a function of $Y_{in}(x_{in})$ for three different choices of the masses. The shaded regions encompass the nearly an order of magnitude variation of the result when assuming the temperature ratio to be constant and a factor of 2 smaller and larger than $T_\chi/T=1$. In the bottom part of the plot the solid lines show the corresponding values of $\lambda$ if the process is instead $\phi\phi \rightarrow \chi\chi$. The huge shift in its value highlights the effect of the suppression of the semi-production channel.

The right panel of Fig.~\ref{fig:illustration} shows the value of the corresponding present-day annihilation cross section as a function of the DM mass for two choices of $Y_{in}(x_{in})$ and for fixed $m_\chi/m_\phi = 10$. Even though in this simple example the resulting cross sections are out of the observational reach, it is clear that the semi-production freeze-in points to the parameter regions that have much greater potential to be probed by indirect detection observations. It is also worth mentioning that if the process was instead $\phi\phi \rightarrow \chi\chi$, then, as expected, there would be no visible dependence on $Y_{in}$, as can be seen from the solid lines coinciding for both choices.

In reality both the temperatures of the DM and the mediator are affected by the decay and annihilation processes in a non-trivial way, \textit{e.g.} the semi-production leads to the self-cooling due to a portion of the kinetic energy being used to produce the heavier state. Therefore, in what follows we study the presented mechanism in more detail, dynamically solving for the temperature evolution. In order to do so we first define a complete example model which also includes self-consistently the initial production process, followed by the period of semi-production. 

\section{The example model}
\label{sec:model}

For a particular realization of the mechanism described in the previous section we consider a Higgs-portal type model with a scalar mediator $\phi$. Such models are extensively studied in the literature and provide a vast phenomenology for collider searches and astroparticle physics (see \textit{e.g.} \cite{Lebedev:2021xey} for a recent review). We take the DM candidate $\chi$ to be a scalar particle,
protected by the $Z_3$ symmetry.
In the absence of any stabilizing symmetry the $\phi$ is expected to mix with the SM Higgs and be unstable, decaying into visible states.\footnote{As before we focus on the regime where $m_{\phi} < 3 m_{\chi}$ to avoid the $\phi$ decay to be the main production channel during the freeze-in process.} The general and renormalizable Lagrangian for such a model is $\mathcal{L}_{int} = \mathcal{L}_{SM} + \mathcal{L}_{\phi-SM}+\mathcal{L}_{DS}$ with
\begin{eqnarray}
\label{eq:Lmodel}
\mathcal{L}_{\phi-SM} &=& A\phi H^\dagger H + \frac{\lambda_{h \phi}}{2} \phi^2  H^\dagger H - \mu_h^2  H^\dagger H +\frac{\lambda_h}{2}( H^\dagger H)^2, \\
\mathcal{L}_{DS} &=& \frac{\mu_\phi^2}{2}\phi^2  +  \frac{\mu_3}{3!}\phi^3 + \frac{\lambda_\phi}{4!}\phi^4 + \mu_\chi^2 \chi^* \chi + \frac{\lambda_\chi}{2} (\chi^* \chi)^2 \\
&&  + \frac{\lambda_1}{3!} \phi\left( \chi^3 + (\chi^*)^3\right) + \frac{\lambda_{2}}{2} \phi^2 (\chi^* \chi) \, ,
\end{eqnarray}
where $H$ stands here for the SM Higgs doublet. To preserve the $Z_3$ symmetry of the vacuum, we require that $\chi$ does not acquire a vacuum expectation value (VEV). For light $\phi$ mediator the branching ratio of the SM Higgs decay to invisible particles is
constrained to be~\cite{Belanger:2013xza} smaller than $0.19$, which
translates to $\lambda_{h\phi}\lesssim 10^{-2}$. Furthermore, we will focus on the region where  couplings $\lambda_{h \phi}$,  $\lambda_1$ and $\lambda_2$ are extremely small, as required by the freeze-in framework. Finally, $\lambda_{\phi}$ and $\lambda_{\chi}$ are free as they are inconsequential for the DM production, as long as they are small enough not to lead to self-thermalization through $2\leftrightarrow 3$ or $2\leftrightarrow 4$ processes.

Before the electroweak phase transition (EWPT) the mass of $\chi$ is simply determined by the parameter $\mu_{\chi}$, while the four massless degrees of freedom of the Higgs doublet  
have the effective masses arising from the temperature corrections. We adopt~\cite{Lebedev:2019ton}
\begin{equation}
m^2_H = -\mu^2_h + \left( \frac{3}{16} g_2^2 + \frac{1}{16} g_1^2 + \frac{1}{4} y_t^2 + \frac{1}{2} \lambda_h \right)T^2\,, 
\end{equation}
where $g_1$, $g_2$ and $y_t$ are the gauge and top Yukawa coupling constants. The mass of the $\phi$ can in principle receive thermal corrections as well, coming from both three- and four-point self-interaction vertices, but in what follows these can be neglected as $\phi$ is significantly underpopulated compared to its equilibrium number density.

Below $T =T_{\rm EW} \sim 160~\GeV$ both $H$ and $\phi$ acquire their VEVs, such that $H = (0, v+h_H)/\sqrt{2} $ (in the unitary gauge) and $\phi = v_{\phi} + \varphi$. The physical scalars appear as the combinations of $h_H$ and $\varphi$
\begin{equation}
    \begin{pmatrix} h \\ \phi \end{pmatrix} = \begin{pmatrix} \cos{\theta} & -\sin{\theta} \\ \sin{\theta} & \cos{\theta} \end{pmatrix} \begin{pmatrix} h_H \\ \varphi \end{pmatrix} \, ,
\end{equation}
where we restore the previously used notation $\phi$ for the mass eigenstate of the new scalar. Assuming that the minimum of the scalar potential corresponds to $\vev{H^\dagger H} = v^2/2$, where $v=246~\GeV$, and that the parameters $A, \mu_{3}$ and $\lambda_{\phi}$ are very small, we obtain the following minimization conditions
\begin{gather}
v_{\phi} \approx -\frac{Av^2}{2m^2_{\phi}} \, , \\
\lambda_h \approx 2 \left(\frac{\mu_h}{v}\right)^2 + \frac{A}{2m^2_{\phi}} \left[1 + \frac{\mu^2_{\phi}}{m^2_{\phi}}\right] \, ,
\end{gather}
where we have used the expression $m^2_{\phi} \approx \mu^2_{\phi} + \lambda_{h\phi}v^2/2$. We are interested in the case when $h$ is essentially the SM Higgs with the mass $m_h \approx 125~\GeV$ and $\phi$ is weakly coupled to it, so that the mixing angle $\theta \ll 1$. Nominally, the mass of $\chi$ also gets a correction and would become $m_{\chi} = \sqrt{\mu^2_{\chi} + \lambda_2 v^2_{\phi}/2}$, but as in our setup both $\lambda_2 \ll 1$ and $v_\phi \ll v$, this shift is negligible.

At very early times, for $T > T_{\rm EW}$ the system is in an unbroken phase where the production of $\phi$ is dominated by the $2\rightarrow 2$ pair-production processes $hh \rightarrow \phi\phi$. After the EWPT, for $T < T_{\rm EW}$, the system moves to a broken phase where $h\rightarrow \phi\phi$ opens up, as long as $m_\phi < m_h/2$, and takes over the role of the most efficient production mode.\footnote{During the EWPT itself there is a short period when the Higgs bosons can convert to $\phi$ through oscillations~\cite{Heeba:2018wtf}, which however does not give rise to an appreciable contribution in the parameter regions we study here.}

For a small mixing angle the mediator $\phi$ is very long-lived, such that it can be considered stable during the freeze-in period. Later it will decay with the lifetime given by the inverse of the decay width $\Gamma_\phi \approx \theta^2 \, \Gamma_{h\to\rm{SM}} (m_\phi)$, where $\Gamma_{h\to\rm{SM}} (m_\phi)$ is the total width of the SM-like Higgs boson with mass $m_\phi$. It follows that the mediator is extremely long-lived at low mass, where only its decays into lepton pairs and photons are kinematically allowed. As we will discuss in the Sec.~\ref{sec:scan}, such long lifetime is severely
constrained by the astrophysical limits and the beam dump limits. Therefore, we will restrict ourselves to $m_\phi > 100$ MeV in the following.

\section{Relic density calculation}
\label{sec:RD}

In this section we introduce the formalism we adopt for determining the amount of produced DM. As we will show, in order to determine the DM relic density in this model one needs to trace the number densities, as well as temperatures, of both $\phi$ and $\chi$.\footnote{For concreteness we will assume that all the interactions conserve CP and that there is no asymmetry between $\chi$ and $\chi^*$, while relaxing this assumption may lead to interesting phenomenological consequences.} This in general leads to a large set of coupled differential equations that, though tractable, are numerically rather expensive. However, in our setup, due to a very weak coupling of $\phi$ and $\chi$ one can simplify the calculation by neglecting the backreaction of the $\chi$ evolution on the temperature of $\phi$, which in turn is then determined solely by the coupling to the SM plasma. Consequently, one can first solve for $(n_\phi, T_\phi)$ and then treat the resulting $T_\phi(T)$ as an input for the remaining evolution of $(n_\chi, n_\phi, T_\chi)$.

\subsection{Boltzmann equations}
\label{sec:BE}

The Boltzmann equation for the evolution of the distribution function $f_{i=\{\chi,\phi\}}$ has the form
\begin{eqnarray}
\label{eq:fBE}
	2E_i \left(\partial_t-H \, p \partial_p \right) f_i (p) =  C\left[f_i \right],
\end{eqnarray}
where the collision term $C\left[f_i \right]$ is described in the next section. Typically the only relevant quantity in the freeze-in scenarios is the produced number density $n_i\equiv g_i\int d^3p \, (2\pi)^{-3} E_i^{-1} \, f_i(p)$, since the dark sector is never populated densely enough for the backreaction processes, nor for the actual momentum distribution of DM, to have any impact. Therefore, only the 0-th moment of the equation \eqref{eq:fBE} is usually considered (but see \cite{Belanger:2020npe} for an exception). In the case of semi-production, however, $\chi$ is also present in the initial state, leading to a dependence of the production rate on its distribution function. 

The most direct approach to tackle this problem is to solve Eq.~\eqref{eq:fBE} completely numerically. This has been done recently in the context of DM freeze-out in \cite{Binder:2017rgn,Binder:2021bmg}, where it has been also shown that solving the system of equations for coupled 0-th and 2-nd moment (cBE) of \eqref{eq:fBE} instead often gives from fairly to very good estimate of the resulting relic density.\footnote{Additionally, the result from cBE is expected to be in fact \textit{very} accurate in the limit of the efficient self-interactions, which redistribute the energy among the DM particles. This limit is especially interesting from the phenomenological perspective as it can lead to the alleviation of the small scale cosmological problems, as we will discuss in Sec. \ref{sec:scan}.} At the same time the structure of the semi-annihilation and semi-production rates is technically more challenging due to the necessity of performing thermal averages on pairs of particles with different temperatures \cite{Kamada:2017gfc,Cai:2018imb,Hektor:2019ote}. Having all this in mind we therefore leave full numerical solution of \eqref{eq:fBE} to future work, while here we focus on the system of cBE.

The derivation of the system of cBE follows exactly \cite{Binder:2017rgn} (see also \cite{vandenAarssen:2012ag} and \cite{Fitzpatrick:2020vba} for a different formulation). Integrating \eqref{eq:fBE} over $g_i\int d^3p/(2\pi)^3/E_i$ and $g_i\int d^3p/(2\pi)^3p^2/E_i^2$
respectively and introducing $x=m_\chi/T$, the yield $Y_i\equiv n_i/s$ and the
'temperature' parameter $y_i \equiv m_i T_i/s^{2/3}$, one gets
\begin{eqnarray}
\frac{Y_i'}{Y_i} &=& \frac{m_i}{x \tilde H}C^0_i\,, \label{Yfinal}\\
\frac{y_i'}{y_i} &=& \frac{m_i }{x \tilde H} C^2_i - \frac{Y_i'}{Y_i} 
+\frac{H}{x\tilde H}
\frac{\langle p^4/E_i^3 \rangle}{3T_i}\,.\label{yfinal}
\end{eqnarray}
Here we also introduce the first two non-vanishing moments of the collision term 
\begin{eqnarray}
C^0_i&\equiv&  \frac{g_i}{m_i n_i} \int \frac{d^3p}{(2\pi)^3E_i}\, C[f_i]\,, \label{eq:C0}\\
C^2_i&\equiv&  \frac{g_i}{3m_i n_i T_i} \int \frac{d^3p}{(2\pi)^3E_i} \frac{p^2}{E_i}\, C[f_i]\, \label{eq:C2}.
\end{eqnarray}
As it can be seen from \eqref{yfinal}, however, additional assumptions are needed to close the Boltzmann hierarchy. In particular, it contains a higher moment
\begin{equation}
\langle p^4/E_i^3 \rangle \equiv \frac{g_i}{n_i} \int \frac{d^3p}{(2\pi)^3}\,\frac{p^4}{E_i^3} f_i(p)
\label{eq:p4E3def} \, ,
\end{equation}
for which, as in \cite{Binder:2017rgn}, we will make an ansatz for both $i=\{\chi,\phi\}$ that
\begin{equation}
f_i(p) = \exp{\left(-(E_i-\mu_i)/T_i\right)} = \big(n_i/n_i^{eq}(T_i) \big) \ f^{eq}_i(p,T_i) \, ,
\label{eq_ansatz}
\end{equation}
where $\mu_i$ is the chemical potential. This ansatz would be exact in the limit of very efficient self-interactions keeping the shape of the distribution functions  close to thermal, albeit with a different temperature. Note that we approximate the equilibrium distribution by the Maxwell-Boltzmann one even though the freeze-in occurs dominantly when $\chi$ and $\phi$ are relativistic. This is justified as both are extremely diluted throughout the whole evolution.\footnote{For a scenario where this is not necessarily the case and one needs to reformulate the cBE to implicitly solve for the chemical potential see \cite{Heeba:2018wtf}.} However, in the numerical computations at high temperatures we treat the SM Higgs in equilibrium as having relativistic distribution, \textit{cf.}~\cite{Arcadi:2019oxh}.

\subsection{The structure of the collision term}
\label{sec:CollTerm}

The collision term $C[f_i]$ can be expressed as a sum of integrals that correspond to different processes in which the particle $i$ participates, \textit{i.e.} $\phi\phi\rightarrow \chi\chi$ pair-production, $\phi\chi\rightarrow \chi\chi$ semi-production, production from the interactions with the SM particles (mainly the Higgs boson) and elastic scatterings:
\begin{equation}
    C[f_i] = C_{\rm \, pair-prod.}[f_i,f_{\chi/\phi}] + C_{\rm \, semi-prod.}[f_i,f_{\chi/\phi}] + C_{\rm \, SM~prod.}[f_i] + C_{\rm el.~scat.}[f_i,f_{\chi/\phi}] \, .
    \label{col_term_proc}
\end{equation}
In principle, each of these terms is different for $\chi$ than for $\phi$, although the structure of the expressions remains similar, while numerical factors and signs can vary. For example, the general expression for the semi-annihilation collision term for the evolution of $\chi$'s distribution function is 
\begin{eqnarray}
\label{semi_colterm}
C_{\rm \, semi-prod.}[f_{\chi}(p_i)] = \frac{1}{g_{\chi}} \int d\Pi_j d\Pi_k d\Pi_l \, (2\pi)^4 \delta^{(4)}(P_i + P_j - P_k - P_l) \; |\mathcal{M}|^2_{\phi\chi^* \leftrightarrow \chi\chi} \nonumber \\ \times \Bigl\{ f_{\phi}(p_k)f_{\chi^*}(p_l) \, \left[1 + f_{\chi}(p_i)\right]\left[1 + f_{\chi}(p_j)\right] - \, f_{\chi}(p_i)f_{\chi}(p_j) \, \left[1 + f_{\phi}(p_k)\right]\left[1 + f_{\chi^*}(p_l)\right] \, \Bigr\}.
\end{eqnarray}
The corresponding term for the evolution of $f_\phi$ has the same structure, but a reverse order of momenta, a different sign and an additional symmetry factor of $1/2$. 

Assuming that the populations of $\chi$ and $\phi$ are very diluted in comparison to their equilibrium values, we can safely neglect the backreaction of $\phi$ on the density of SM plasma and omit all of the contributions to the collision term that are $\mathcal{O}(f_{\chi,\phi}^2)$. Thus, for calculation purposes is convenient to further rewrite all the expressions like Eq.~\eqref{semi_colterm} in terms of specific combinations of distribution functions.
For the system of cBE we do not require the full unintegrated collision term as above, but rather only its $0$-th and $2$-nd moments. The $0$-th moment, Eq.~\eqref{eq:C0}, governs the rate of the number density evolution and is basically an integral of the collision term over the phase-space of the particle under consideration. The elastic scattering part contribution to the collision term vanishes after this integration and does not affect the density evolution, as expected. Considering Eq.~\eqref{semi_colterm} as an example and using the ansatz from Eq.~\eqref{eq_ansatz} and the approach described in the previous paragraph, one can notice that the $0$-th moment can be formulated simply in terms of the velocity-averaged cross sections for various $2 \rightarrow 2$ processes (or in terms of the width for decays), \textit{e.g.}
\begin{equation}
    C_{\phi\phi \rightarrow \chi\chi^* } [f_{\phi}(p), f_{\phi}(k)] \propto n^2_{\phi} \, \vev{\sigma v}_{\phi\phi \rightarrow \chi\chi^* } \, .
    \label{term_sigma_v}
\end{equation}

The $2$-nd moment, Eq.~\eqref{eq:C2}, governs the rate of the temperature evolution and is considerably more complicated. The additional $p^2/E$ factor makes the integration more involved and the $2$-nd moment terms cannot be in general expressed as simply as in Eq.~\eqref{term_sigma_v}. In particular in the case of semi-annihilations such terms require additional numerical integrations and hence the solution of the cBE in full generality takes noticeably more time. Similarly, the $2$-nd moment of the elastic scatterings term also connects particles with different temperatures leading to the same complications. Note, that unlike in the freeze-out case here one cannot justify the expansion in the Fokker-Planck type collision term~\cite{Bringmann:2006mu,Binder:2016pnr}. However, we have checked with explicit integration of the scattering term that for the interaction strengths considered, the momentum transfer rate is too small to have any impact. 

Therefore, after including all the decay, pair-annihilation and semi-annihilation processes, in the absence of an asymmetry between $\chi$ and $\chi^*$ one arrives at the system of cBE that we solve numerically:
\vspace{0.2cm}
\begin{align}
& \frac{Y_\phi'}{Y_\phi} = \frac{m_\phi}{x \tilde H}\, \left(\, ^0_{\phi}C_{\underline{\phi}\phi \rightarrow \chi\chi^*} + \, ^0_{\phi}C_{\underline{\phi}\chi^* \rightarrow \chi\chi} + \, ^0_{\phi}C_{\underline{\phi}\chi \rightarrow \chi^*\chi^*} + \, ^0_{\phi}C_{h \rightarrow \underline{\phi}\phi} + \, ^0_{\phi}C_{hh \rightarrow \underline{\phi}\phi} \right), \label{Yphi}\\
& \frac{y_\phi'}{y_\phi} = \frac{m_\phi }{x \tilde H} \left( \, ^2_{\phi}C_{h \rightarrow \underline{\phi}\phi} + \, ^2_{\phi}C_{hh \rightarrow \underline{\phi}\phi} \right) - \frac{Y_\phi'}{Y_\phi} 
+\frac{H}{x\tilde H}
 \frac{\langle p^4/E_\phi^3\rangle}{3T_\phi}\,,\label{yphi} \\
& \frac{Y_{\rm DM}'}{Y_{\rm DM}} = \frac{m_\chi}{x \tilde H}\, \left( \,  ^0_{\chi}C_{\phi\chi^* \rightarrow \underline{\chi}\chi} + \, ^0_{\chi}C_{\phi\underline{\chi} \rightarrow \chi^*\chi^*} + \, ^0_{\chi}C_{\phi\phi \rightarrow \underline{\chi}\chi^*}  \right), \label{YDM}\\
& \frac{y_\chi'}{y_\chi} = \frac{m_\chi }{x \tilde H} \left(\, ^2_{\chi}C_{\phi\underline{\chi} \rightarrow \chi^*\chi^*} + \, ^2_{\chi}C_{\phi\chi^* \rightarrow \underline{\chi}\chi} + \, ^2_{\chi}C_{\phi\phi \rightarrow \underline{\chi}\chi^*} \right) - \frac{Y_{\rm DM}'}{Y_{\rm DM}} 
+\frac{H}{x\tilde H}
\frac{\langle p^4/E_\chi^3\rangle}{3T_\chi}\,, \label{y} 
\end{align}
\vspace{0.2cm} 

\noindent where $Y_{\rm DM}=Y_\chi+Y_{\chi^*}$ and the complete set of expressions entering the moments of the collision term
can be found in the Appendix \ref{append}. 

As mentioned above, due to a very weak coupling between the states in the dark sector, for the temperature evolution of $\phi$ we can consider only the interactions with the SM plasma, while neglecting the backreaction of $\chi$. That is, we first solve Eq.~\eqref{Yphi} coupled only with~\eqref{yphi} to determine the relation $T_\phi(T)$. Next, we use it as an input for the system of coupled Eqns.~\eqref{Yphi}, \eqref{YDM} and \eqref{y}.

\subsection{Results for benchmark points}

\begin{figure}[t]
\centering
\includegraphics[scale=0.53]{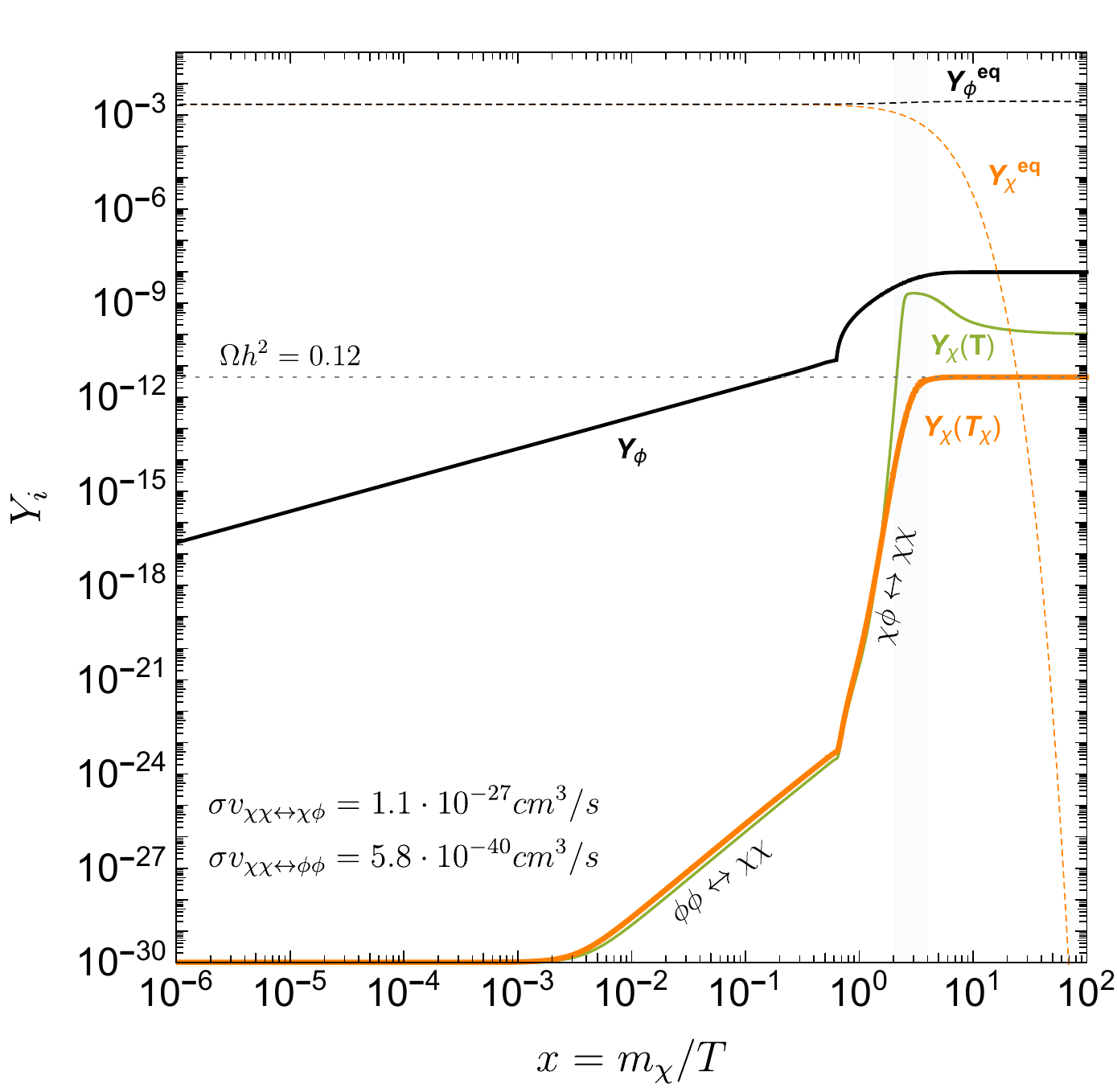}
\includegraphics[scale=0.521]{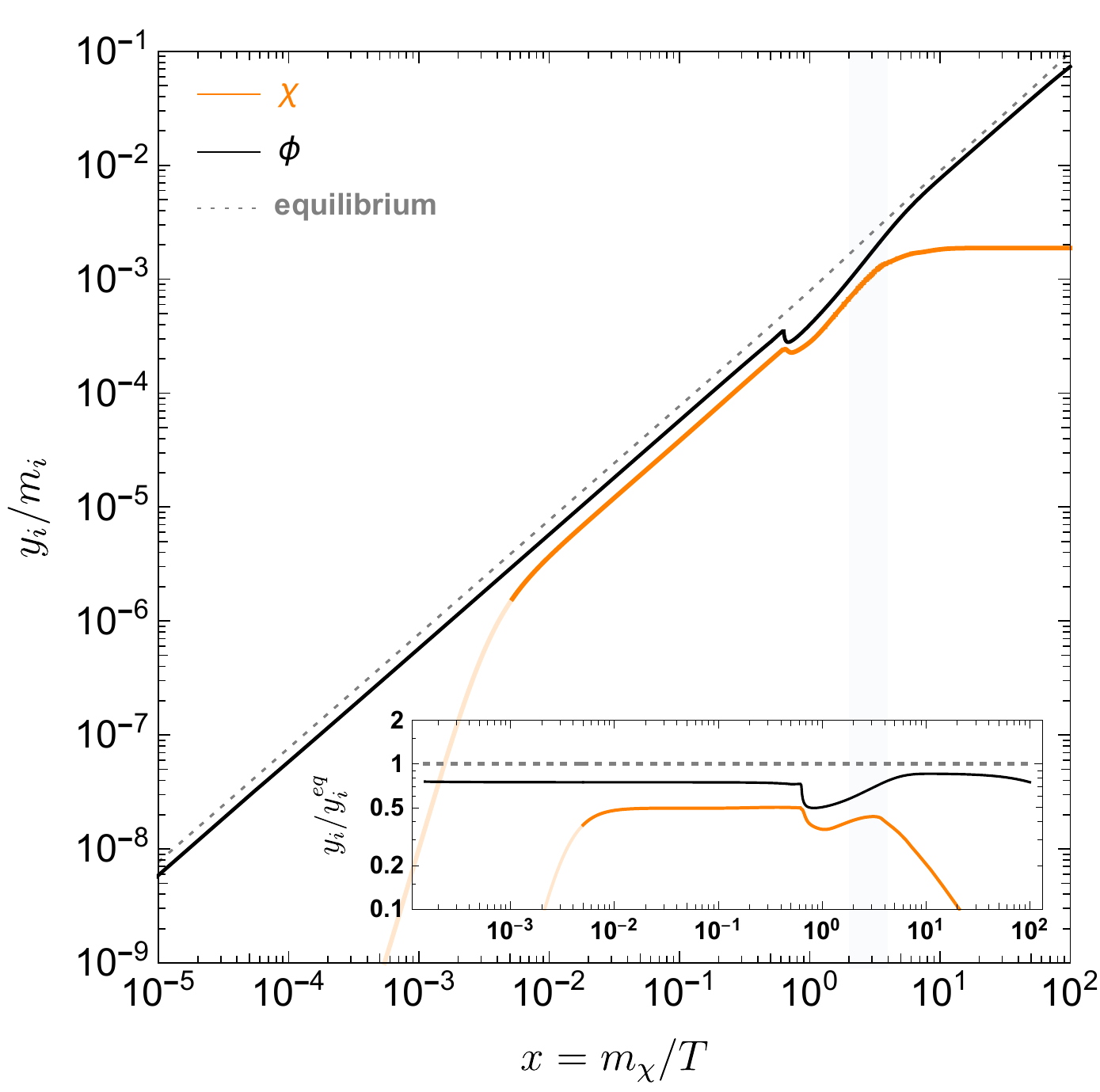}
\caption{The evolution of $\chi$ and $\phi$ number densities \textit{(left)} and temperatures \textit{(right)} for a benchmark point with $m_\chi=100$ GeV, $\mu_\phi=1$ GeV, $\lambda_1=1.1\times 10^{-2}$, $\lambda_2=10^{-8}$, $\lambda_{h \phi}=6 \times 10^{-11}$ and $\theta=10^{-5}$, which makes semi-production the dominant  process. \textit{Left}: the solid lines show the evolution of $Y_\phi$ (black), $Y_\chi$ (orange) and $Y_\chi(T_\chi=T)$ (green), while the dotted correspond to their equilibrium values. The process specified below the orange line highlights the dominant production channel in a given regime. \textit{Right}: in the main plot the solid lines show the evolution of $y_\phi$ (black) and $y_\chi$ (orange), while in the inset plot their ratio to the equilibrium value is shown. The part of the orange line with lighter shading shows the regime where no significant population has yet been produced. In all of the plots the light gray shaded region depicts the time window in which the production of $\chi$ is most efficient.
}
\label{fig:BM_semi}
\end{figure}

\begin{figure}[t]
\centering
\includegraphics[scale=0.53]{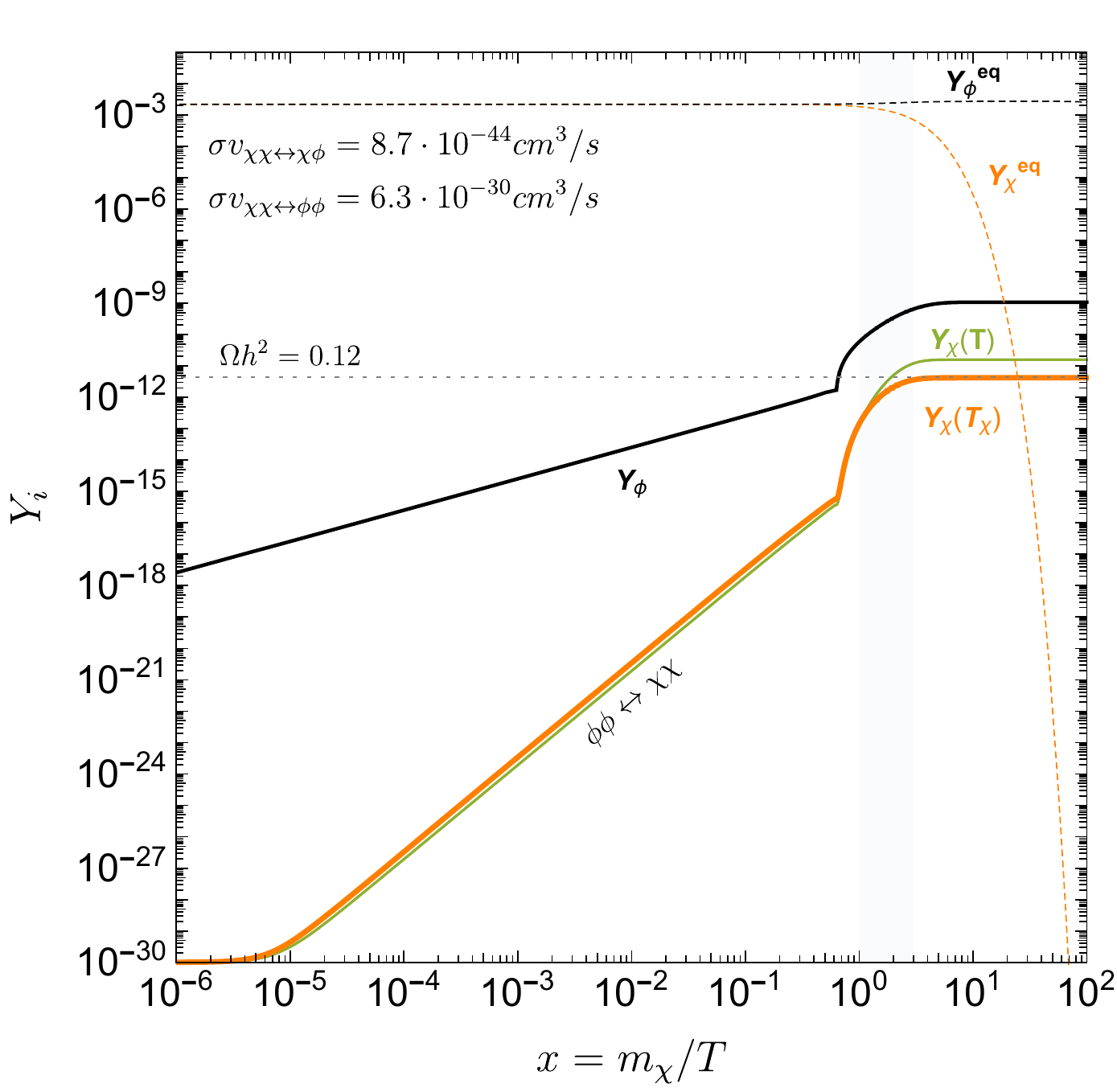}
\includegraphics[scale=0.521]{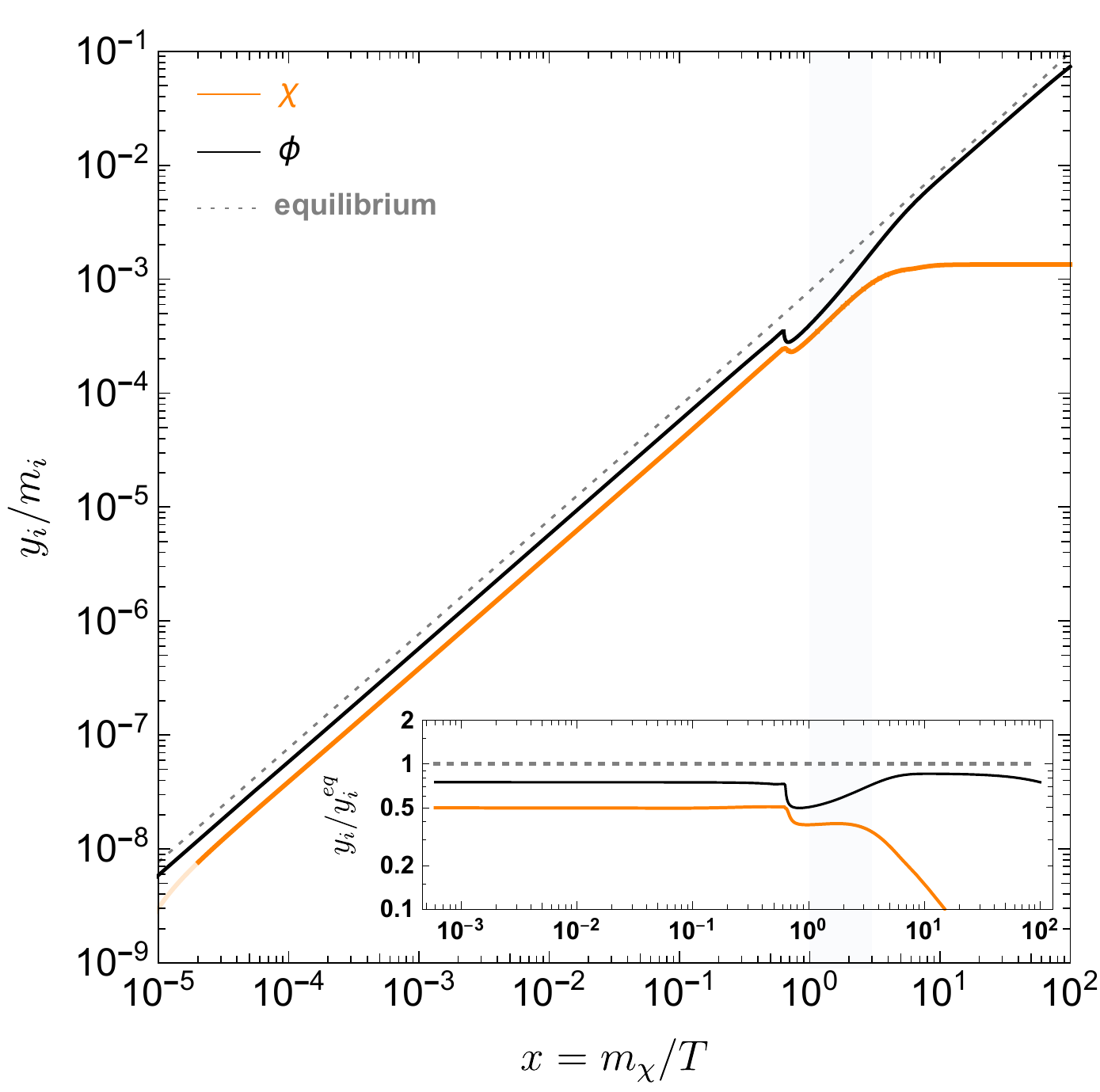}
\caption{The same as in Fig.~\ref{fig:BM_semi}, but for $m_\chi=100$ GeV, $\mu_\phi=1$ GeV, $\lambda_1=10^{-10}$, $\lambda_2=10^{-3}$, $\lambda_{h \phi}=2 \times 10^{-11}$ and $\theta=10^{-5}$, which makes pair-production the dominant process.}
\label{fig:BM_2to2}
\end{figure}

In the next section we present the results for our model's parameter space scan, but first let us start with a discussion of the thermal history of $\chi$ and $\phi$ using as examples two representative benchmark points. These are shown in Figs.~\ref{fig:BM_semi} and \ref{fig:BM_2to2} for two choices of the parameters that lead to the DM creation mode, which is dominated by semi-production and pair-production respectively. In both figures the left panel gives the evolution of the yields $Y_\chi$ and $Y_\phi$, while the right panel shows the departure of the $y$ parameters from the corresponding equilibrium ones illustrating the magnitude of the departure from kinetic equilibrium with the SM plasma.

Looking at the left panels of the figures one can distinguish four phases of the evolution:
\begin{description}
    \item{I)} At very high temperatures one is in an  unbroken SU(2) phase and $\phi$ is being produced from $2 \rightarrow 2$ processes, while the production of $\chi$, being dependent upon $Y_\phi$, is still negligible.
    \item{II)} After some non-negligible population of $\phi$ is built up, the $\phi\phi \rightarrow \chi\chi$ process starts to slowly create DM particles. This phase is essential to give the initial seed for the semi-production to occur.
    \item{III)} After the EWPT the $h\rightarrow \phi\phi$ decay is switched on, which leads to an accelerated production of $\phi$ and indirectly $\chi$ as well. The growing populations of both particles exponentially boost the increase of $Y_{\chi}$ via semi-production, given that the corresponding coupling is considerably large.\footnote{It should be noted that when the coupling between $\chi$ and $\phi$ is very strong, it can happen that they will equilibrate with each other \cite{Bernal:2020gzm}, even though both are relatively diluted. This can also be a result of self-interactions~\cite{Arcadi:2019oxh},  however this requires couplings that are much larger than the ones discussed in our scenario.} 
    \item{IV)} When the temperature drops further all the interaction processes freeze in and both number densities reach a plateau. Not shown in the plot, but worth mentioning is that later on $\phi$ will eventually decay through the mixing with the SM Higgs and thus will not contribute to the DM relic abundance.
\end{description}

Turning now to the right panels of Figs.~\ref{fig:BM_semi} and \ref{fig:BM_2to2} showing the temperature change one can observe that for most of the evolution, both $\phi$ and $\chi$ have temperatures of the same order of magnitude as $T$. This is of course expected, as at early times the $2\rightarrow 2$ processes tend to produce $\phi$ and $\chi$ with energies of order $\sim T$ and since $T\gg m_\chi,\, m_\phi$, they redshift in the same way as the SM plasma. Note that in reality the $T_\phi$ and $T_\chi$ are somewhat lower than $T$, 
which is related to the fact that  in the relativistic regime particles in the low energy part of the distribution have a higher probability of interacting.
After the EWPT point is reached, the $h\rightarrow \phi\phi$ introduces a visible drop in the $\phi$ temperature, and then indirectly $T_\chi$ as well, followed by a slow rise later on. This is because at the onset of EWPT, when $h$ is still relativistic, the decay leads to $E_{\phi} \sim T/2$, while later the same process gives $E_\phi \sim m_h/2\gg T$. Finally, for the evolution of $T_\chi$ at $x\sim $ few the production freezes-in and $\chi$ completely decouples from the rest of the states, both chemically and kinetically.

The light gray region on both plots is shown to guide the eye to the region of $x$ in which the most of the DM production takes place, to help visualise how much the actual temperatures are different then from the assumption of kinetic equilibrium. The part of the orange line with lighter shading shows the regime where no significant population has yet been built up and the value of $y$ there is dependent on the initial condition, which later becomes washed out after more DM is produced.

Additionally, the green solid line shows the result for $Y_\chi$ under the simplifying assumption that both $\chi$ and $\phi$ remain in kinetic equilibrium throughout the whole production period. For both benchmark points this overshoots the more accurate result based on solving the cBE system from factor of a few, when pair-production dominates, to more than on order of magnitude, when it is the semi-production that is more efficient. Two effects contribute to such a significant overestimate when tracing only the number density. Firstly, neglecting the change in the temperature of $\phi$ leads to an incorrect thermal average for both the $\phi\phi \rightarrow \chi\chi$ and $\chi\phi \rightarrow \chi\chi$ processes, impact of which is the greater the smaller $m_\phi$ is compared to $m_\chi$. In particular, since the actual $T_\phi$ is smaller than $T$, this thermal average overestimates the production rate. Secondly, with strong semi-production comes the dependence on $T_\chi$ as well. Together with an exponential growth period it leads to a much larger change with respect to the naive expectation. 

Therefore, one can appreciate the necessity of solving for not only the number density, but also for the temperature in order to arrive at an accurate estimate of the relic abundance. It should be mentioned, however, that from the phenomenological perspective, at least for the model at hand, this exponential dependence means only logarithmic change of the coupling $\lambda_1$ is sufficient to mitigate the error made by assuming the kinetic equilibrium. Nevertheless, it is worth stressing that the departure from the kinetic equilibrium can influence the production process even more strongly if, unlike in the case of our example model, the interactions are substantially velocity-dependent.

\section{Scan results and phenomenology}
\label{sec:scan}

When $\phi$ is in the chemical equilibrium with the SM plasma the semi-production mechanism leads to the enhanced values of the coupling compared to the typical freeze-in, but as illustrated in Sec.~\ref{sec:illustration}, the resulting cross sections are still somewhat below the sensitivity of the near-future indirect searches. However, combined with the idea of the sequential freeze-in \cite{Belanger:2020npe}, \textit{i.e.} simply when $\phi$ is coupled very weakly to the SM states such that it also undergoes the freeze-in, the resulting signals can be much stronger.

In order to study the phenomenological implications of the proposed mechanism we performed a numerical scan within the specified model to find the points that satisfy the relic density constraint. We scanned over the following ranges with the logarithmic sampling for the five parameters: $m_\chi \in \left[10^{-1}, 10^3\right]$ GeV, $m_\phi \in \left[10^{-1}, 50\right]$ GeV, $\lambda_1 \in \left[10^{-10}, 1\right]$, $\lambda_2 \in \left[10^{-10}, 1\right]$, $\lambda_{h\phi} \in \left[10^{-15}, 10^{-8}\right]$ and $\theta \in \left[10^{-8}, 10^{-2}\right]$. The ranges were chosen to cover a wide range of masses for $\chi$ and $\phi$, with the conditions that $m_\chi>m_\phi$ and $m_\phi<m_h/2$. As for the couplings, we restricted $\lambda_{h\phi}$ to be concentrated in the freeze-in region of $\phi$ and the mixing angle to be small enough ($\theta \lesssim 10^{-2}$) not to be in conflict with the collider measurements \cite{OConnell:2006rsp}.

\begin{figure}[t]
\centering
\includegraphics[scale=0.8]{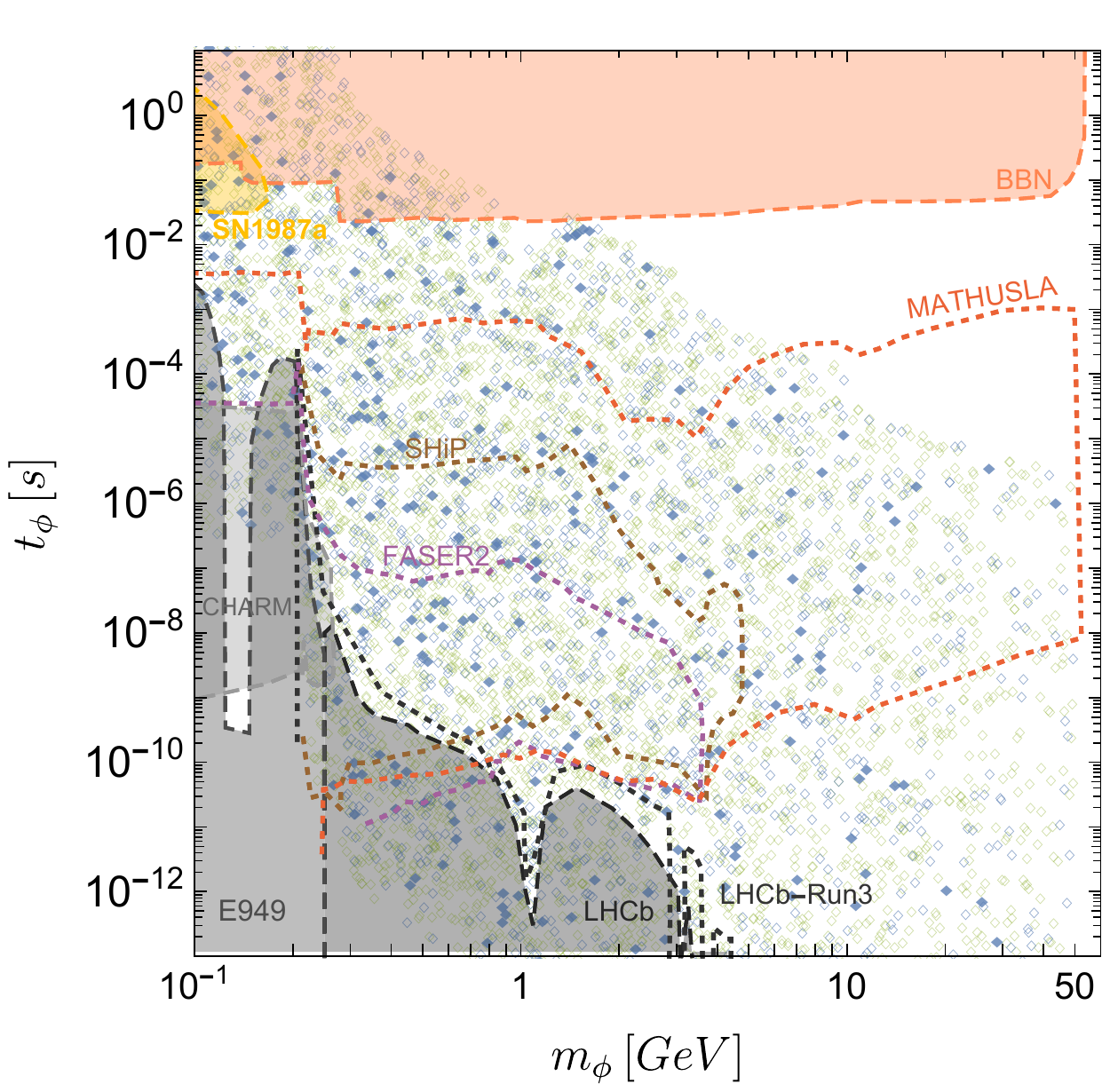}
\caption{The results of the scan showing points with $\Omega h^2= 0.12\pm 0.012$ projected onto the $(m_\phi,t_\phi)$ plane. The blue (green) colors indicate points with the dominant semi-annihilation $\sigma v_{\chi\chi \rightarrow\chi\phi} > \sigma v_{\chi\chi \rightarrow\, \phi\phi}$ (pair-annihilation $\sigma v_{\chi\chi \rightarrow \chi\phi} < \sigma v_{\chi\chi \rightarrow \phi\phi}$) process. The superimposed constraints on the mediator $\phi$ are shown as shaded regions: recasted \cite{Winkler:2018qyg} results from CHARM \cite{Bergsma:1985qz}, E949 \cite{Artamonov:2008qb}
and LHCb \cite{Aaij:2016qsm,Aaij:2015tna} (gray); the limits from BBN \cite{Fradette:2017sdd} (orange) and SN1987a \cite{Krnjaic:2015mbs} (yellow). Also we show the prospects for the future experiments with dashed lines: FASER \cite{Ariga:2018uku} (violet), SHiP \cite{Anelli:2015pba} (blue) and MATHUSLA \cite{Alpigiani:2020tva} (red) and LHCb-Run3 (black) adopted from~\cite{Lanfranchi:2020crw}.
}
\label{fig:scanphi}
\end{figure}

\begin{figure}[t]
\centering
\includegraphics[scale=0.59]{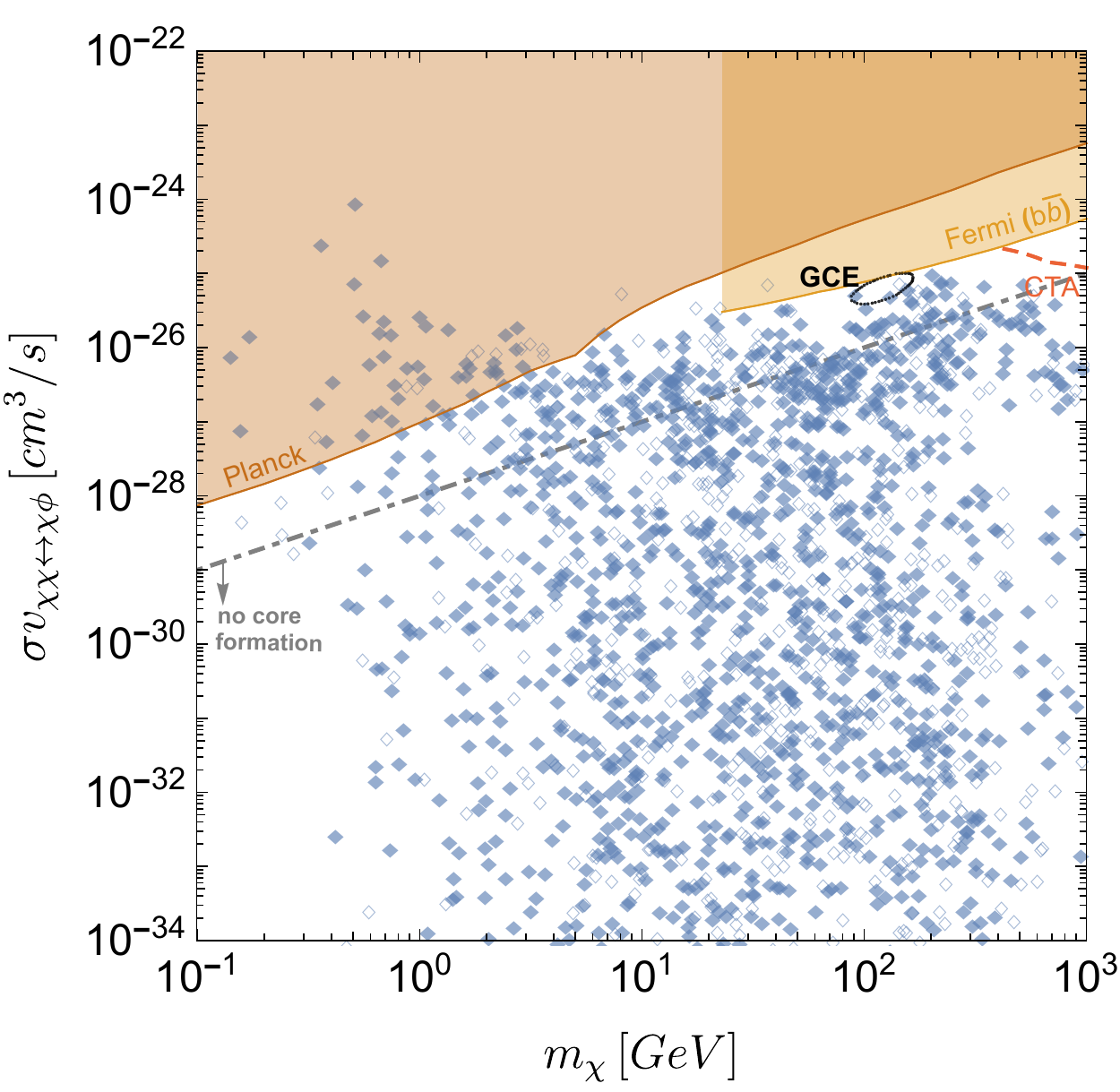}
\includegraphics[scale=0.59]{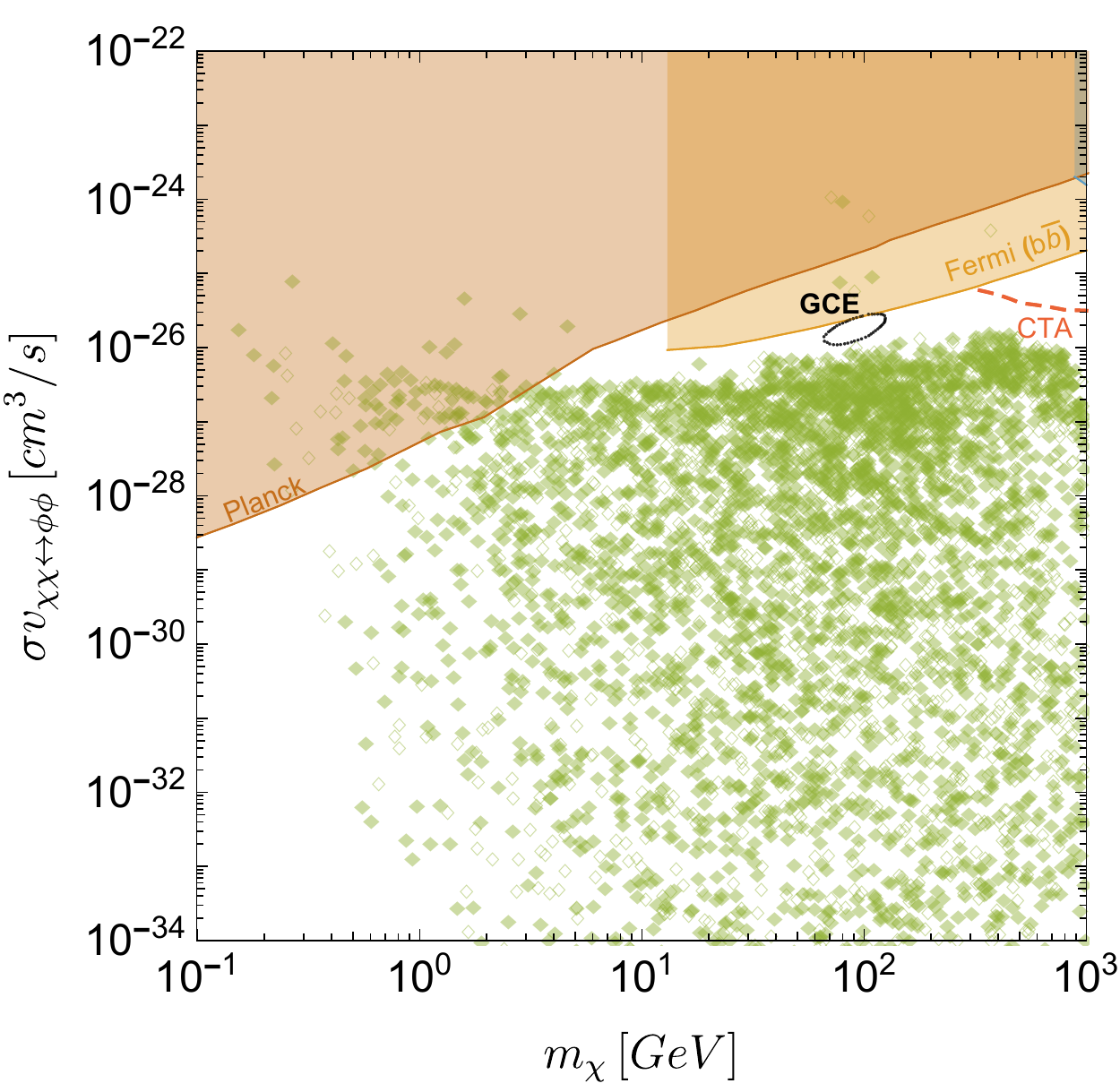}
\caption{The results of the scan showing the points that correspond to $\Omega h^2= 0.12\pm 0.012$ with the superimposed indirect detection limits. Filled diamonds indicate the points that are within the reach of the future searches of the mediator $\phi$, \textit{cf.} Fig.~\ref{fig:scanphi} while the empty ones are beyond these prospects. The existing limits from Planck measurements of the CMB \cite{Aghanim:2018eyx} (light brown) and Fermi-LAT+MAGIC combined analysis of the gamma-ray observations of dSphs \cite{Ahnen:2016qkx} (orange) are shown as shaded regions, while the dashed red line gives the projected sensitivity of the CTA for the $b\bar b$ channel \cite{Acharyya:2020sbj}.
\textit{Left}: points with dominant semi-annihilation $\sigma v_{\chi\chi \rightarrow\chi\phi} > \sigma v_{\chi\chi \rightarrow\, \phi\phi}$ projected onto the $(m_\chi,\sigma v_{\chi\chi \rightarrow\chi\phi})$ plane.  The black ellipse shows the best fit region for the GC excess recasted from \cite{Cuoco:2016jqt}. The dot-dashed gray line indicates the semi-annihilation cross section below which there is no appreciable core formation in the dSph by the self-heating of DM \cite{Chu:2018nki}. 
\textit{Right}: points with dominant pair-annihilation $\sigma v_{\chi\chi \rightarrow\chi\phi} < \sigma v_{\chi\chi \rightarrow\, \phi\phi}$ projected onto the $(m_\chi,\sigma v_{\chi\chi \rightarrow\phi\phi})$ plane.
}
\label{fig:scanID}
\end{figure}

The results of the scan are shown in Figs.~\ref{fig:scanphi} and~\ref{fig:scanID}. First in Fig.~\ref{fig:scanphi} all the points found are projected onto the mass vs. lifetime of $\phi$ plane and confronted with the limits from the light boson searches. The points are represented as blue (green) diamonds for the regime of semi-annihilation domination $\sigma v_{\chi\chi \rightarrow\chi\phi} > \sigma v_{\chi\chi \rightarrow\, \phi\phi}$ (pair-annihilation $\sigma v_{\chi\chi \rightarrow \chi\phi} < \sigma v_{\chi\chi \rightarrow \phi\phi}$). 
Additionally, some of the blue diamonds are filled, signifying the points that have the present-day semi-annihilation cross section large enough to potentially affect the core formation in the dwarf galaxies, following the results of \cite{Chu:2018nki}.

The shaded regions show the recast of the existing limits, while the dashed lines show the prospects for the upcoming upgrades and planned experiments. There is a visible expected correlation between the $m_\phi$ and the lifetime $t_\phi$, but apart from that, one can see that the most of the allowed parameter region in this plane is accessible for the studied model. 

In Fig.~\ref{fig:scanID} the same points are separated into ones with dominant semi-annihilation (left panel, blue points) and pair-annihilation (right panel, green points) and projected onto the plane of dominant cross section vs. $m_\chi$.\footnote{A small subset of the points will have both processes contributing at the same level. For these one should actually combine them both and only then compare with the constraints. However, this does not change the overall results of the scan and would unnecessarily complicate the discussion.}
Filled diamonds indicate the points that are within the reach of the future searches for the mediator $\phi$, \textit{cf.} Fig.~\ref{fig:scanphi}, while the empty ones are beyond the prospects. The points that are in conflict with the above measurements are discarded. 

In both panels we show the constraints from the CMB measurements by the Planck collaboration \cite{Aghanim:2018eyx}, recasted using \cite{Slatyer:2015jla}, the limits from the joint analysis of the gamma-ray flux from dwarf spheroidal galaxies (dSphs) performed by Fermi-LAT and MAGIC collaborations \cite{Ahnen:2016qkx} and the prospects for CTA 
\cite{Acharyya:2020sbj}.
All these limits need to be taken as an illustration of the actual constraints that by necessity apply differently to different points, \textit{e.g.} due to varying $m_\phi$ that changes the final photon spectra. For the purpose of demonstration we took the constraints on the DM annihilations to $b\bar{b}$ and shifted them accordingly such that the $\sigma v$ limit for a given value of $m_{\chi}$ corresponds to the aforementioned limits for the DM particle that has the same mass as half of the energy of $\phi$. We take into account the corresponding normalization for the yield and the DM number density and we assume that the difference in the gamma-ray spectra between these two cases can be neglected. For pair-annihilation $E_{\phi} = m_{\chi}$ and for semi-annihilation $E_{\phi} = (m^2_{\phi} + 3m^2_{\chi})/4m_{\chi}$ (assuming both processes happen at rest). In the broad mass range of interest $\phi$ decays predominantly to quarks through the mixing with the SM Higgs, so the $b\bar{b}$ mode is a satisfactory approximation.

In the same fashion we recast the $2\sigma$ best-fit parameter region for the explanation of the GC gamma-ray excess with a quite similar scalar-singlet model \cite{Cuoco:2016jqt}. These regions are marked with black ellipses in Fig.~\ref{fig:scanID}. 
This is done in order to emphasize that the freeze-in production can indeed be used to construct models explaining the GC excess (see also \cite{Heikinheimo:2018duk}). Note, that although we use a rather simplified approach to obtain the best-fit region of the parameters, our estimations are in a reasonable agreement with the results of a dedicated fit to the semi-annihilation signal described in \cite{Arcadi:2017vis}. 

Finally, in the semi-annihilation case we show the line separating the cross sections that might lead to a sufficient heat transfer in the inner parts of the dwarf galaxies to contribute to the formation of the core~\cite{Chu:2018nki}. Above this thin gray line, depending on the efficiency of the energy redistribution among the DM particles, one might affect the core formation.\footnote{However, see \cite{Kamada:2019wjo} for a more recent analysis, which finds this solution to be in tension when confronted also with the observed central mass deficit
of field dwarf and low surface brightness spiral galaxies.}

From these results one can infer that the best search strategy for the model at hand is through the searches of the light mediator: a significant portion of the studied parameter space is going to be covered by the beam dump experiments. On the indirect detection frontier typically the predicted (semi-)annihilation cross sections are well below the values probed by current and near-future experiments. However, some of the model realizations do give the signals that are potentially detectable. Moreover, one should stress that even though there are not so many points found in the scan that are within the current limits, the high cross-section region requires more fine-tuning and is harder to find through a random scan. Therefore, the existence of even relatively few points like this in the scan results does prove that these regions exist even in the simple example model at hand.

\section{Conclusions}
\label{sec:conclusions}

Dark matter produced through freeze-in is an attractive alternative to its more studied thermal freeze-out cousin which is typically significantly more challenging to detect, especially in the indirect searches. In this work we introduced a novel thermal freeze-in production mechanism which we called \textit{semi-production}, that leads to DM candidates having much larger coupling values than in the usual freeze-in, and consequently, present day annihilation cross section potentially giving observable signals in the indirect searches.

In order to calculate the relic density in an accurate way we employed a set of coupled Boltzmann equation for the number density and temperature of DM interacting with the out-of-equilibrium mediator. We show that neglecting 
the temperature evolution can lead to a result for the relic abundance that is off by more than an order of magnitude, especially in the parameter space regions with dominant semi-production. 

Furthermore, we have performed a numerical scan of the parameter space of an example model which shows on a quantitative level that one can indeed explain the current abundance of DM through the freeze-in mechanism and at the same time predict large enough (semi-)annihilation cross section to be phenomenologically relevant. In particular, it can reach the values needed for fitting the GC excess, as well as semi-annihilation cross sections that can help to alleviate the core-cusp problem in dSphs.

Even though most of the parameter space of the model studied here predicts the signals that are too weak to be observed even in a not-so-near future, we would like to emphasize that the proposed mechanism is quite general and can be used in other models as well. 
The fact that the semi-production freeze-in relies on the only prerequisite that the $\phi\chi\rightarrow \chi\chi$  process dominates over all other interaction channels and that it leads to a quite intriguing phenomenology constitutes a promising basis for further investigations into model building employing this mechanism.

\paragraph{Note added:} During the final stages of completion of this paper, an article  \cite{Bringmann:2021lyf} appeared being first to propose in essence the same mechanism as studied here. Although, as far as the production mechanism is concerned, there are no crucial qualitative differences between our works, nevertheless, our study goes quite beyond the discussion of \cite{Bringmann:2021lyf} by calculating the temperature evolution, considering the out-of-equilibrium mediator and investigating indirect detection signals for a specific model realization.

\acknowledgments

We would like to thank Dimitrios Karamitros for involvement and very helpful discussions during the initial stage of the project. This work is supported by the National Science Centre, Poland, research grant No. 2018/31/D/ST2/00813.

\appendix

\section{The moments of the collision term}
\label{append}

Here we derive the expressions for the moments of the collision terms introduced in Eqs.~\eqref{eq:C0} and \eqref{eq:C2}. Note that presented results are not completely general, but follow the approximations as described in Sec.~\ref{sec:CollTerm}.   

It is convenient to express these moments for each type of particle as a sum of the integrals $^N_{i}C_{A \rightarrow B}$ that correspond to different processes $A \rightarrow B$, in which a particle $i$ is involved and $N$ denotes the order of the moment of the collision term. The particle $i$ under consideration always corresponds to the momentum $p$ and its position in the reaction is underlined. We are going to use the following notation

\begin{equation}
    d\Pi^{(n)}_i = d\Pi_p \cdot ... \cdot d\Pi_{(n)} \, (2\pi)^4 \delta^{(4)}\lrb{ \sum_A p_A - \sum_B p_B  } \, ,
\end{equation}
where $d\Pi_{p}\equiv \frac{d^3p}{(2\pi)^3 2E_p}$, $n$ represents the number of states in the corresponding reaction $A \rightarrow B$ and the delta-function determines the 4-momentum conservation. For simplicity we use the same notation for each amplitude squared $|\mathcal{M}|^2$, but one has to note that it always corresponds to the reaction specified in the notation of the respective contribution to the collision term. Moreover, in the model that we study $|\mathcal{M}|^2$ is constant for all of the processes, so in the explicit expressions we are going to put it as a factor in front of the integrals.
For the distribution functions of $\chi$ and $\phi$ everywhere below we substitute the ansatz from Eq.~\eqref{eq_ansatz} with the corresponding temperatures.

\subsubsection*{The 0-th moment}

Here we list all the relevant contributions needed for both evolution of the abundance of $\chi$ and $\phi$ in a form suitable for numerical computations.

\paragraph{Pair-production.}

These contributions contain the same type of distribution functions among the particles in both initial and final states 
\begin{equation}
\quad ^0_{\chi}C_{\phi\phi \rightarrow \underline{\chi}\chi^*} = \frac{1}{2m_{\chi} n_{\chi}}  \int d\Pi^{(4)}_{\chi} \Big[ f_{\phi}(q) f_{\phi}(r) - f_{\chi}(p) f_{\chi^*}(k) \Big] |\mathcal{M}|^2 \,.
\end{equation}
Thus, they can be expressed in the usual form using the thermally averaged cross section
\begin{equation}
\quad ^0_{\chi}C_{\phi\phi \rightarrow \underline{\chi}\chi^*} = \frac{1}{m_{\chi} n_{\chi}}  \Biggr[ n^2_{\phi} \, \vev{\sigma v}_{\phi\phi \rightarrow \chi\chi^*}(T_{\phi})  - n^2_{\chi} \, \vev{\sigma v}_{\chi\chi^* \rightarrow \phi\phi}(T_{\chi}) \Biggr].
\end{equation}
The factor $1/2$ takes into account the symmetry between the identical particles from the perspective of the Boltzmann equation.\footnote{The Boltzmann equation keeps track of the states with a certain momentum, while the other momenta are integrated over. Integrating over a set of $n$ identical states yields a factor of $1/n!$. For example, in the process $\underline{\phi}\phi \rightarrow \chi\chi^*$ the BE keeps track of the $\phi(p)$, while $\phi(k)$ is integrated over ($\chi$ and $\chi^*$ are not identical). In the process $\phi\phi \rightarrow \underline{\chi}\chi^*$ the BE keeps track of $\chi(p)$, while both $\phi$ states are integrated over, which yields a symmetry factor of $1/2$. A useful discussion on these symmetry factors can be found in the notes inside~\cite{DK}.
}

The pair-annihilation term for $\phi$ can be expressed through the respective term for the evolution of $\chi$ 
\begin{equation}
\quad ^0_{\phi}C_{\underline{\phi}\phi \rightarrow \chi\chi^*} = - 2 \, \frac{m_{\chi} n_{\chi}}{m_{\phi} n_{\phi}} \;  ^0_{\chi}C_{\phi\phi \rightarrow \underline{\chi}\chi^*}\,.
\end{equation}
Note that the factor of 2 here compensates the symmetry factor from the respective term for $\chi$.

\paragraph{Semi-production.} 

These contributions can still be formulated in terms of the velocity-averaged cross sections for the corresponding processes, however, due to different distribution functions cannot be cast in a form of one dimensional integral. In general we have
\begin{equation}
\quad ^0_{\chi}C_{\phi\chi^* \rightarrow \underline{\chi}\chi} = \frac{1}{m_{\chi} n_{\chi}} \int d\Pi^{(4)}_{\chi} \left[ f_{\phi}(q) f_{\chi}(r) - f_{\chi}(p) f_{\chi}(k) \right] \, |\mathcal{M}|^2\,.
\end{equation}
The inverse, semi-annihilation, term can be calculated exactly as in the case of pair-annihilation above, while the more complicated semi-production one can be expressed as
\begin{multline}
^0_{\chi}C_{\phi\chi^* \rightarrow \underline{\chi}\chi} [f_{\phi}(q), f_{\chi}(r)] = \frac{|\mathcal{M}|^2}{(2\pi)^5 \cdot 16 m_{\chi}  \, \eq{n}{\chi} }  \left(\frac{n_{\phi}}{\eq{n}{\phi}}\right) \Biggr(\frac{n_{\chi}^*}{n_{\chi}}\Biggr) \times \\  \times \int^{+\infty}_{m_{\phi}} \; dE_q \; \eq{f}{\phi} (q)  \int^{+\infty}_{E^{\rm min}_r (q)} \; dE_r \; \eq{f}{\chi} (r)
\int^{s_{\rm max}(E_q,E_r)}_{s_{\rm min}(E_q,E_r)} ds \; 
\tilde p (s),
\label{C0_1}
\end{multline}
where $\tilde p (s) = \sqrt{s - 4m^2_{\chi}} / 2\sqrt{s}$ is the momentum of $\chi$ in the center-of-mass (CM) frame,  $s_{\text{max/min}} (E_q,E_r) = m^2_{\phi} + m^2_{\chi} + 2E_q E_r \pm 2qr$ denote the maximal and the minimal values of the $s$-invariant for the states with momenta $q$ and $r$ and $E^{\rm min}_r (q)$ is the energy limit such that the reaction with the given energy of $E_q$ is kinematically possible.
The integral over $s$ can be done analytically, so the computation of such terms for the processes with a constant squared amplitudes require double numerical integration.

To get the term for a conjugated process one has to switch $n_{\chi}$ and $n_{\chi^*}$ in the expression above
\begin{equation}
\quad ^0_{\chi}C_{\phi\underline{\chi} \rightarrow \chi^*\chi^*} = - \frac{1}{2} \; ^0_{\chi}C_{\phi\chi^* \rightarrow \underline{\chi}\chi} \Big\rvert_{\chi \rightarrow \chi^*}\,.
\end{equation}
For $\phi$ the semi-annihilation contributions can be also expressed through the terms above
\begin{align}
& ^0_{\phi}C_{\underline{\phi}\chi^* \rightarrow \chi\chi} = - \frac{1}{2} \, \frac{m_{\chi} n_{\chi}}{m_{\phi} n_{\phi}} \;  ^0_{\chi}C_{\phi\chi^* \rightarrow \underline{\chi}\chi}\,,\\
& ^0_{\phi}C_{\underline{\phi}\chi \rightarrow \chi^*\chi^*} = \; ^0_{\phi}C_{\underline{\phi}\chi^* \rightarrow \chi\chi}\,.
\end{align}

\paragraph{Production from the SM plasma.}

Interactions between $\phi$ and the SM Higgs boson $h$ give the following contributions to the 0-th moment of the collision term
\begin{equation}
\quad ^0_{\phi}C_{h \rightarrow \underline{\phi}\phi} = \frac{1}{m_{\phi} n_{\phi}} \int d\Pi^{(3)}_{\phi} \eq{f}{h} (q) \Big( 1 + 2f_{\phi} (p) \Big) |\mathcal{M}|^2\,.
\end{equation}
The first term here can be included in full detail, however in practice it is sufficient to use a simpler, analytical expression, where one adapts the Maxwell-Boltzmann approximation for the distribution function of $h$ (note that this process is active only after EWPT where $T$ is of order $m_h$ or lower):
\begin{equation}
^0_{\phi}C_{h \rightarrow \underline{\phi}\phi} [f_h (q)] \approx \frac{  m_h \, T_{\rm SM} \, |\mathcal{M}|^2}{ (2\pi)^3 \cdot 4  m_{\phi} n_{\phi}} \sqrt{1 - \frac{4m^2_{\phi}}{m^2_h}} K_1 \left( \frac{m_h}{T_{\rm SM}} \right), 
\end{equation}
where $K_1$ is the modified Bessel function of the second kind. The second term containing $f_\phi(p)$ is more complicated
\begin{equation}
^0_{\phi}C_{h \rightarrow \underline{\phi}\phi} [f_h (q), f_{\phi} (p)] = \frac{T_{\phi} |\mathcal{M}|^2 }{ (2\pi)^3 \cdot  m_{\phi} \eq{n}{\phi}} \int^{+\infty}_{m_h} dE_q \, \eq{f}{h} (q) \sinh{\left( \frac{q}{2T_{\phi}}\right)} \exp{\left( -\frac{E_q}{2T_{\phi}}\right)},
\end{equation}
where we assume for the simplicity of the expression that $m_h \gg m_{\phi}$.

Finally, the $hh$ pair-production gives the following contribution
\begin{equation}
\quad ^0_{\phi}C_{hh \rightarrow \underline{\phi}\phi} = \frac{1}{2m_{\phi} n_{\phi}} \int d\Pi^{(4)}_{\phi} \Big[ \eq{f}{h} (q) \eq{f}{h} (r)  - f_{\phi} (p) f_{\phi} (k) \Big] |\mathcal{M}|^2\,,
\end{equation}
which again boils down to the same structure of integration as for the pair-production above. Note, however, that for the SM Higgs we do keep the Fermi-Dirac form of the distribution function.

The other contributions from the Higgs-mediated interactions with the SM plasma are strongly suppressed by the small  mixing angle $\theta$.

\subsubsection*{The 2-nd moment}

As we state in the text, the calculation of the 2-nd moment terms is more elaborate. If the factor $p^2/E$ corresponds to the momentum of a product particle, the integration over $s$ (see Eq.~\eqref{C0_1}) cannot be performed analytically, so several terms require triple numerical integrations.\footnote{Including the terms with a distribution function in the integration over the final states (such as the terms of $\mathcal{O}(f_{\chi,\phi}^2)$) would lead to an even larger number of numerical integrations.}

\paragraph{Semi-production.} 

Let us start with the contribution of semi-production to the 2-nd moment term for the thermal evolution of $\chi$:
\begin{equation}
\quad ^2_{\chi}C_{\phi\underline{\chi} \rightarrow \chi^*\chi^*} = \frac{1}{6m_{\chi} n_{\chi}T_{\chi}} \int d\Pi^{(4)}_{\chi} \, \frac{p^2}{E} \, \left[ f_{\chi^*}(q) f_{\chi^*}(r) - f_{\chi}(p) f_{\phi}(k) \right] |\mathcal{M}|^2 \,.
\end{equation}
The first term in the above expression is of higher order $\mathcal{O}(f_\chi^2)$ and thus we neglect it, just as other terms of this order below.
The second term can be expressed as follows
\begin{multline}
^2_{\chi}C_{\phi\underline{\chi} \rightarrow \chi^*\chi^*} [f_{\chi}(p), f_{\phi}(k)] = \frac{ |\mathcal{M}|^2}{(2\pi)^5 \cdot 96 m_{\chi} T_{\chi}  \, \eq{n}{\chi} }  \left(\frac{n_{\phi}}{\eq{n}{\phi}}\right) \int^{+\infty}_{m_{\phi}} \; dE_k \; \eq{f}{\phi} (k) \times \\  \times \int^{+\infty}_{E^{\rm min}_p (k)} \; dE_p \; \eq{f}{\chi} (p) \left( E_p - \frac{m^2_{\chi}}{E_p}\right)
\int^{s_{\rm max}(E_k,E_p)}_{s_{\rm min}(E_k,E_p)} ds \; \sqrt{1 - \frac{4m^2_{\chi}}{s}} ,
\end{multline}
where we use the same notations as in Eq.~\eqref{C0_1} and again the integral over $s$ can be easily done analytically, but just leads to a longer expression. The conjugated term now cannot be simply expressed through the previous ones:
\begin{equation}
\quad ^2_{\chi}C_{\phi\chi^* \rightarrow \underline{\chi}\chi} = \frac{1}{3m_{\chi} n_{\chi}T_{\chi}} \int d\Pi^{(4)}_{\chi} \, \frac{p^2}{E} \, \left[ f_{\phi}(q) f_{\chi}(r) - f_{\chi}(p) f_{\chi}(k) \right]  |\mathcal{M}|^2\,.
\end{equation}
Although the second term in this expression is not difficult to include it is formally of higher order and we skip it for consistency, while the first can be expressed as:
\begin{multline}
^2_{\chi}C_{\phi\chi^* \rightarrow \underline{\chi}\chi} [f_{\chi}(r), f_{\phi}(q)] = \frac{ |\mathcal{M}|^2}{(2\pi)^5 \cdot 24 m_{\chi} T_{\chi}  \, \eq{n}{\chi} }  \left(\frac{n_{\phi}}{\eq{n}{\phi}}\right) \Biggr(\frac{n_{\chi^*}}{n_{\chi}}\Biggr)  \int^{+\infty}_{m_{\phi}} \; dE_q \; \eq{f}{\phi} (q) \times \\  \times \int^{+\infty}_{E^{\rm min}_r (q)} \; dE_r \; \eq{f}{\chi} (r) 
\int^{s_{\rm max}(E_r,E_q)}_{s_{\rm min}(E_r,E_q)} ds \; \Biggr[ \frac{\tilde p \, \tilde E_p (E_q + E_r) }{s} - \frac{m^2_{\chi} \; \text{ArcTanh}(\tilde p v / \tilde E_p)}{\sqrt{(E_q + E_r)^2 - s}} \Biggr] \, ,
\label{C2:triple}
\end{multline}
where $\tilde p(s)$ and $\tilde E_p(s)$ are the momentum and the energy of the product $\chi$ in the center-of-mass frame and $v = \sqrt{1 - s/(E_q + E_r)^2}$ is the velocity of the CM frame.

\paragraph{Pair-production.} 

For the pair-annihilation we only keep the $f^2_{\phi}$ contribution

\begin{equation}
\quad ^2_{\chi}C_{\phi\phi \rightarrow \underline{\chi}\chi^*} = \frac{1}{6m_{\chi} n_{\chi}T_{\chi}} \int d\Pi^{(4)}_{\chi} \, \frac{p^2}{E} \,  f_{\phi}(q) f_{\phi}(r) \; |\mathcal{M}|^2\,.
\end{equation}
The resulting expression is analogous to Eq.~\eqref{C2:triple} with the proper adjustments for the factors, distribution functions and kinematic variables. 

\paragraph{Production from the SM plasma.}
 
As explained in the main text for the temperature evolution of $\phi$ we can completely neglect the backreaction from $\chi$ and concentrate on the processes involving only the SM Higgs boson:
\begin{multline}
\quad ^2_{\phi}C_{h \rightarrow \underline{\phi}\phi} = \frac{1}{3m_{\phi} n_{\phi}T_{\phi}} \int d\Pi^{(3)}_{\phi} \, \frac{p^2}{E} \,  \eq{f}{h}(q) \Big( 1 + f_{\phi} (p)\Big)\;  |\mathcal{M}|^2 \; \approx \; \frac{|\mathcal{M}|^2}{(2\pi)^3 \cdot 6m_{\phi}n_{\phi}T_{\phi}} \times \\ 
\times \int^{+ \infty}_{m_h} dE_q \, q \; \eq{f}{h} (q) \sqrt{1 - \frac{4m^2_{\phi}}{m^2_{h}}} \Biggr[ \frac{E_q}{4} - \frac{m^2_{\phi} \, \text{ArcTanh}\left(q \sqrt{1 - 4m^2_{\phi}/m^2_h} \, / E_q  \right)}{q \sqrt{1 - 4m^2_{\phi}/m^2_h}} \Biggr].
\end{multline}
Another relevant term comes from the $hh$ pair-annihilation 
\begin{equation}
\quad ^2_{\phi}C_{hh \rightarrow \underline{\phi}\phi} = \frac{1}{6m_{\phi} n_{\phi}T_{\phi}} \int d\Pi^{(4)}_{\phi} \, \frac{p^2}{E} \,  \eq{f}{h} (q) \eq{f}{h} (r) \;  |\mathcal{M}|^2 \, ,
\end{equation}
for which an expanded expression can be adapted from Eq.~\eqref{C2:triple}.


\bibliographystyle{JHEP}  

\providecommand{\href}[2]{#2}\begingroup\raggedright\endgroup

\end{document}